\documentclass[journal]{IEEEtran}
\usepackage{lineno,hyperref}
\modulolinenumbers[5]

\usepackage{tikz}
\usepackage{enumerate}
\usepackage{amsfonts}
\usepackage{amssymb}
\usepackage{amsmath}
\usepackage{bbm}
\usepackage{graphicx}
\usepackage[utf8]{inputenc}
\usepackage[english]{babel}
\usepackage{algorithm}
 \usepackage{algorithmic}
\usepackage{color}

\DeclareMathOperator{\Tr}{Tr}

\usepackage{xspace}
\usepackage{intmacros}
\usepackage{subfigure}

\newcounter{examplecounter}
\renewcommand{\theexamplecounter}{\arabic{examplecounter}}

\makeatletter
\@ifclasslater{IEEEtran}{2012/12/26}
 {\newcommand{\killproofname}{\unskip\nopunct}}
 {\newcommand{\killproofname}[1]{\unskip\aftergroup\ignorespaces\ignorespaces}}
\makeatother

\title{Diffusion LMS for Distributed Estimation over Wireless Networks with Inter-Node Interference Perturbation}
\author{Mohammadjavad Mirzazadeh Moallem and 
        Mehdi~Korki,~\IEEEmembership{Member,~IEEE}
\thanks{M. M. Moallem (contact author) is with Babol Noshirvani University of Technology, Babol, Iran. M. Korki is with School of Science, Computing and Engineering Technologies, Swinburne University of Technology, Melbourne, Australia. Email: mkorki@swin.edu.au.
}}

\begin{document}
\bstctlcite{IEEEexample:BSTcontrol}
\maketitle
\sloppy 

%





\begin{abstract}
In this paper, we investigate the diffusion least mean square (DLMS) algorithm over fading channel, where in addition to channel noise and path-loss the inter-node-interference (INI) among neighboring nodes of a host node is also taken into account. We also analyze the mean-square convergence
behavior of DLMS algorithm, under such condition. In addition, based on an upper bound of the derived network MSD, an optimization problem is defined to find an optimal combination strategy. Furthermore, the adaptive version of the proposed combination strategy is presented. Simulation results corroborate the  theoretical findings and indicate the superiority of the proposed combination methods over some previously reported algorithms.
\end{abstract}




%

\section{Introduction}\label{sec_1}
\begin{figure}[htb]
\centering

\tikzset{every picture/.style={line width=0.75pt}} 

\begin{tikzpicture}[x=0.75pt,y=0.75pt,yscale=-1,xscale=1]

\draw  [fill={rgb, 255:red, 0; green, 0; blue, 0 }  ,fill opacity=1 ] (151.99,109.66) .. controls (151.99,106.94) and (153.97,104.72) .. (156.42,104.72) .. controls (158.87,104.72) and (160.86,106.94) .. (160.86,109.66) .. controls (160.86,112.39) and (158.87,114.6) .. (156.42,114.6) .. controls (153.97,114.6) and (151.99,112.39) .. (151.99,109.66) -- cycle ;
\draw  [line width=1.5] [line join = round][line cap = round] (161.44,109.14) .. controls (185.97,109.14) and (207.92,109.22) .. (232.71,109.22) ;
\draw  [line width=1.5]  (234.41,110.91) .. controls (234.41,106.82) and (237.38,103.5) .. (241.06,103.5) .. controls (244.73,103.5) and (247.71,106.82) .. (247.71,110.91) .. controls (247.71,115) and (244.73,118.32) .. (241.06,118.32) .. controls (237.38,118.32) and (234.41,115) .. (234.41,110.91) -- cycle ;
\draw  [fill={rgb, 255:red, 0; green, 0; blue, 0 }  ,fill opacity=1 ] (206.8,105.96) -- (213.45,109.66) -- (206.8,113.37) -- (210.13,109.66) -- cycle ;
\draw [line width=1.5]    (247.71,110.91) -- (308.45,110.21) ;
\draw [shift={(312.45,110.17)}, rotate = 179.34] [fill={rgb, 255:red, 0; green, 0; blue, 0 }  ][line width=0.08]  [draw opacity=0] (10.45,-5.02) -- (0,0) -- (10.45,5.02) -- cycle    ;
\draw [line width=1.5]    (241.06,79.55) -- (241.06,99.5) ;
\draw [shift={(241.06,103.5)}, rotate = 270] [fill={rgb, 255:red, 0; green, 0; blue, 0 }  ][line width=0.08]  [draw opacity=0] (10.45,-5.02) -- (0,0) -- (10.45,5.02) -- cycle    ;
\draw  [fill={rgb, 255:red, 0; green, 0; blue, 0 }  ,fill opacity=1 ] (312.45,110.17) .. controls (312.45,107.44) and (314.44,105.23) .. (316.89,105.23) .. controls (319.33,105.23) and (321.32,107.44) .. (321.32,110.17) .. controls (321.32,112.9) and (319.33,115.11) .. (316.89,115.11) .. controls (314.44,115.11) and (312.45,112.9) .. (312.45,110.17) -- cycle ;
\draw  [fill={rgb, 255:red, 0; green, 0; blue, 0 }  ,fill opacity=1 ] (174.23,208.46) .. controls (174.94,205.83) and (177.44,204.22) .. (179.8,204.86) .. controls (182.17,205.5) and (183.5,208.16) .. (182.79,210.79) .. controls (182.07,213.42) and (179.58,215.03) .. (177.21,214.39) .. controls (174.85,213.75) and (173.51,211.09) .. (174.23,208.46) -- cycle ;
\draw [color={rgb, 255:red, 208; green, 2; blue, 27 }  ,draw opacity=1 ][line width=1.5]  [dash pattern={on 5.63pt off 4.5pt}]  (182.45,206.55) -- (310.2,117.27) ;
\draw [shift={(313.48,114.98)}, rotate = 145.05] [fill={rgb, 255:red, 208; green, 2; blue, 27 }  ,fill opacity=1 ][line width=0.08]  [draw opacity=0] (10.45,-5.02) -- (0,0) -- (10.45,5.02) -- cycle    ;
\draw  [fill={rgb, 255:red, 0; green, 0; blue, 0 }  ,fill opacity=1 ] (254.6,269.01) .. controls (256.34,266.91) and (259.28,266.47) .. (261.16,268.03) .. controls (263.05,269.59) and (263.17,272.56) .. (261.43,274.66) .. controls (259.69,276.77) and (256.75,277.2) .. (254.86,275.64) .. controls (252.98,274.08) and (252.86,271.11) .. (254.6,269.01) -- cycle ;
\draw [color={rgb, 255:red, 208; green, 2; blue, 27 }  ,draw opacity=1 ][line width=1.5]  [dash pattern={on 5.63pt off 4.5pt}]  (259.62,266.76) -- (315.62,124.09) ;
\draw [shift={(317.08,120.37)}, rotate = 111.43] [fill={rgb, 255:red, 208; green, 2; blue, 27 }  ,fill opacity=1 ][line width=0.08]  [draw opacity=0] (10.45,-5.02) -- (0,0) -- (10.45,5.02) -- cycle    ;
\draw [line width=1.5]    (262.39,341.23) -- (325,341.48) ;
\draw [shift={(329,341.5)}, rotate = 180.23] [fill={rgb, 255:red, 0; green, 0; blue, 0 }  ][line width=0.08]  [draw opacity=0] (10.45,-5.02) -- (0,0) -- (10.45,5.02) -- cycle    ;
\draw [line width=1.5]    (240.97,106.6) -- (241.07,114.84) ;
\draw [line width=1.5]    (244.94,110.44) -- (236.9,110.75) ;
\draw  [fill={rgb, 255:red, 0; green, 0; blue, 0 }  ,fill opacity=1 ] (374.79,270.44) .. controls (374.68,267.71) and (376.58,265.43) .. (379.02,265.33) .. controls (381.47,265.23) and (383.54,267.36) .. (383.65,270.08) .. controls (383.76,272.81) and (381.87,275.1) .. (379.42,275.2) .. controls (376.98,275.29) and (374.9,273.17) .. (374.79,270.44) -- cycle ;
\draw [color={rgb, 255:red, 208; green, 2; blue, 27 }  ,draw opacity=1 ][line width=1.5]  [dash pattern={on 5.63pt off 4.5pt}]  (377.03,265.41) -- (323.33,121.86) ;
\draw [shift={(321.93,118.12)}, rotate = 69.49] [fill={rgb, 255:red, 208; green, 2; blue, 27 }  ,fill opacity=1 ][line width=0.08]  [draw opacity=0] (10.45,-5.02) -- (0,0) -- (10.45,5.02) -- cycle    ;
\draw [color={rgb, 255:red, 208; green, 2; blue, 27 }  ,draw opacity=1 ][line width=1.5]  [dash pattern={on 5.63pt off 4.5pt}]  (263.39,362.23) -- (326,362.48) ;
\draw [shift={(330,362.5)}, rotate = 180.23] [fill={rgb, 255:red, 208; green, 2; blue, 27 }  ,fill opacity=1 ][line width=0.08]  [draw opacity=0] (10.45,-5.02) -- (0,0) -- (10.45,5.02) -- cycle    ;

\draw (185.18,85.34) node [anchor=north west][inner sep=0.75pt]  [font=\normalsize]  {$\mathit{h_{1}{}_{k}}( i)$};
\draw (356.01,263.15) node [anchor=north west][inner sep=0.75pt]  [font=\Huge,rotate=-175.73]  {$\textcolor[rgb]{0.82,0.01,0.11}{\cdots }$};
\draw (223.88,56.9) node [anchor=north west][inner sep=0.75pt]  [font=\normalsize,rotate=-0.7]  {$\boldsymbol{n}_{1}{}_{k}{}_{,}{}_{i}$};
\draw (136.38,85.03) node [anchor=north west][inner sep=0.75pt]  [font=\small,rotate=-77.97]  {$Neighbor\ 1$};
\draw (295.43,84.45) node [anchor=north west][inner sep=0.75pt]  [font=\small,rotate=-0.58]  {$Node\ k$};
\draw (349.39,297.43) node [anchor=north west][inner sep=0.75pt]  [font=\small,rotate=-336.06]  {$Neighbor\ m_{k}$};
\draw (375.59,230.91) node [anchor=north west][inner sep=0.75pt]  [font=\normalsize,rotate=-249.49]  {$\textcolor[rgb]{0.82,0.01,0.11}{h_{m}{}_{_{k} k}( i)}$};
\draw (200.92,159.7) node [anchor=north west][inner sep=0.75pt]  [font=\normalsize,color={rgb, 255:red, 208; green, 2; blue, 27 }  ,opacity=1 ,rotate=-326.49]  {$h_{2}{}_{k}( i)$};
\draw (148.79,187.29) node [anchor=north west][inner sep=0.75pt]  [font=\small,rotate=-58.52]  {$Neighbor\ 2$};
\draw (247.16,226.62) node [anchor=north west][inner sep=0.75pt]  [font=\normalsize,color={rgb, 255:red, 208; green, 2; blue, 27 }  ,opacity=1 ,rotate=-290.43]  {$h_{3}{}_{k}( i)$};
\draw (219.9,273.44) node [anchor=north west][inner sep=0.75pt]  [font=\small,rotate=-20.04]  {$Neighbor\ 3$};
\draw (162.71,333.13) node [anchor=north west][inner sep=0.75pt]  [font=\small,rotate=-359.7]  {$Desired\ Signal$};
\draw (139.71,355.13) node [anchor=north west][inner sep=0.75pt]  [font=\small,color={rgb, 255:red, 208; green, 2; blue, 27 }  ,opacity=1 ,rotate=-359.7]  {$Interfering\ Signal$};

\end{tikzpicture}
\caption{Inter-node interference (INI) at node $k$} 
\label{fig:INI}
\end{figure}
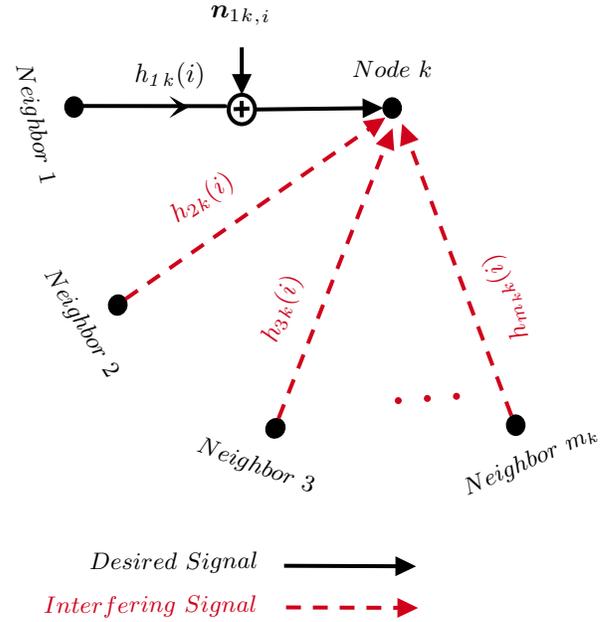

%
%
%
%
Distributed networks consist of numerous interconnected nodes which continuously learn and adapt from measurements to estimate an unknown vector in a distributed manner. Distributed estimation is a widely accepted method in different applications, especially in wireless sensor networks (WSNs), where scalability, robustness, and low energy consumption are essential \cite{ref_5}. These methods benefit from localized in-network processing and inter-node data exchange to solve an estimation problem in a cooperative and online manner. In this paper, among incremental \cite{ref_1}, consensus \cite{ref_2}, and diffusion \cite{ref_5,ref_3,ref_6} strategies, we focus on diffusion-based algorithms for the estimation of an unknown vector parameter. Recently, several efforts have been done to overcome the challenges encountered in diffusion-based wireless networks. One of the most important challenges is analyzing the performance of the algorithm when the wireless links between the nodes are non-ideal, i.e., they include some perturbations such as fading and additive noise. There is, however, another perturbation called inter-node interference (INI) which should be taken into account in practice. This is because, in diffusion-based algorithms, the neighboring nodes share some intermediate signals among each other simultaneously to estimate the unknown vector \cite{ref_27}. In addition, all nodes are equipped with radio modules working in the same frequency, for example, Wi-Fi frequency band. Therefore, the received signal at each node, e.g., node $k$, is affected by the superposition of transmitted data from the transmitting nodes (see Figure \ref{fig:INI}). In other words, an intermediate signal transmitted by a neighboring node $l$ of node $k$ will be affected by the signals transmitted by other neighboring nodes (instead of only node $l$) of node $k$. On the other hand, using different transmitters and receivers with different frequency bands is not cost-effective or feasible in networks with such limited resources. As a result, the received signal is perturbed by INI.

In this paper, we aim to investigate the performance of the diffusion least-mean-square (DLMS) algorithm in the presence of fading, additive noise, and INI. Towards that end, we demonstrate that the algorithm converges in both mean and mean-square senses in the presence of fading, additive noise, and INI, with some assumptions. Furthermore, a left-stochastic matrix will be proposed to govern the combination of data delivered at node $k$. The entries of this matrix are derived by solving a minimization problem via convex optimization frameworks.

Several diffusion algorithms have been proposed over ideal communication channel \cite{ref_5}, \cite{ref_3}\nocite{ref_6}--\cite{ref_22}. In addition, there exists some work such as \cite{ref_8}, \cite{ref_9}, and \cite{ref_23} on the analysis of diffusion algorithms under an additive noise with Gaussian distribution. However, fading and path loss also contaminate the received signals, and they can degrade the performance of the algorithm. Hence, in \cite{ref_11}, \cite{ref_20}, and \cite{ref_21} Abdolee et al. studied the DLMS algorithm over fading channels. In particular, in \cite{ref_21}, they analyze DLMS algorithm in a wireless network with dynamic topology and with channel estimation using pilot signals.

Various combination methods have also been proposed such as Laplacian \cite{ref_24} and Maximum Degree \cite{ref_26} strategies. However, these strategies are inappropriate for imperfect channel scenarios and they lead to the performance degradation of the diffusion algorithm. Hence, in \cite{ref_23} and \cite{ref_12} two combination weights strategies have been proposed, which utilize the channel state information (CSI), to enhance the performance of the DLMS algorithm, in the presence of channel impairments. Nevertheless, such studies lack the theoretical analyses and verification of the simulation and theoretical findings. We derive a left-stochastic combination weights matrix by formulating an optimization problem using an upper bound of the network mean-square deviation (MSD). Then a closed-form optimal solution to the problem is computed. These combination weights are adaptive, i.e., changing with the adaptive topology of the network, in the presence of fading, additive noise, and INI.
In this paper, we extend adapt-then-combine (ATC) version of DLMS algorithm to a wireless network with a time-varying topology where data is shared among nodes subject to INI, in addition to fading and channel noise impairments. The contributions of this paper are summarized as follows: first, we consider, for the first time in the literature, a new model for the intermediate received signal at each node $k$, which includes INI along with fading and noise. Based on the proposed model, we then analyze the  mean and mean-square convergence of the DLMS algorithm, under such conditions. We also proposed an adaptive combination weights matrix using a convex optimization framework. Our simulation results reveal a good fit with theoretical findings and also the superiority of the proposed combination method over some state-of-the-art algorithms.

{\it Notations}: $\mathbb{C}$ denotes the field of  complex numbers. Scalars are denoted by lower-case letters, and vectors and matrices respectively by lower- and upper-case boldface letters. The transpose and complex conjugate-transpose are denoted by $(\cdot)^{\mathrm{T}}$ and $(\cdot)^{*}$, respectively. $\mathbb{E}\{\cdot\}$ represents expectation. $\boldsymbol{I}_{M}$ denotes an $M \times M$ identity matrix. $\otimes$ denotes the Kronecker product operation. $\text{diag}\left\{ \cdot\right\}$ represents a diagonal matrix with its arguments. $\text{col}\left\{ \cdot\right\}$ denotes an enlarged column vector structured by stacking its columns on top of each other. $\mathbf{1}_{M}$ and $\mathbf{0}_{M}$ are the column vector of length $M$ with all entries being one and zero, respectively. $\mathbf{1}_{M \times N}$ is an $M \times N$ matrix with all entries being one. $\left \| \boldsymbol{x}\right \|$ denotes the Euclidean norm of its vector argument. Let $\boldsymbol{x}=\operatorname{col}\left\{\boldsymbol{x}_{1}, \boldsymbol{x}_{2}, \ldots, \boldsymbol{x}_{N}\right\}$ denote an $N \times 1$ block column vector whose individual entries are of size $M \times 1$ each. Hence, the block maximum norm of $\boldsymbol{x}$ is denoted by $\|\boldsymbol{x}\|_{b, \infty}$ and is defined as $\|\boldsymbol{x}\|_{b, \infty} \triangleq \max _{1 \leq k \leq N}\left\|\boldsymbol{x}_{k}\right\|$. Correspondingly, the induced block maximum norm of an arbitrary $N \times N$ block matrix $\boldsymbol{\mathcal{A}}$, whose individual block entries are of size $M \times M$ each, is defined as $\|\boldsymbol{\mathcal{A}}\|_{b, \infty} \triangleq \max _{\boldsymbol{x} \neq 0} \frac{\|\boldsymbol{\mathcal{A}} \boldsymbol{x}\|_{b, \infty}}{\|\boldsymbol{x}\|_{b, \infty}}$. We define the eigenvalue set of the square matrix $\boldsymbol{X}$ as $\{\lambda(\boldsymbol{X})\}$, with $\lambda_{\max }(\boldsymbol{X})$ denoting the maximum eigenvalue. The spectral radius of the square matrix $\boldsymbol{X}$ is denoted by $\rho(\boldsymbol{X}) \triangleq \max \{|\lambda(\boldsymbol{X})|\}$. $\|\boldsymbol{x} \|^2_{\boldsymbol{\Sigma}}$ denotes the weighted vector norm, i.e., $\| \boldsymbol{x} \|^2_{\boldsymbol{\Sigma}}=\boldsymbol{x}^*\boldsymbol{\Sigma} \boldsymbol{x}$ for any Hermitian $\boldsymbol{\Sigma}>0$. $\operatorname{vec}(\boldsymbol{X})$ vectorizes matrix $\boldsymbol{X}$ and stacks its columns on top of each other.


\section{Signal Model}\label{sec_2}
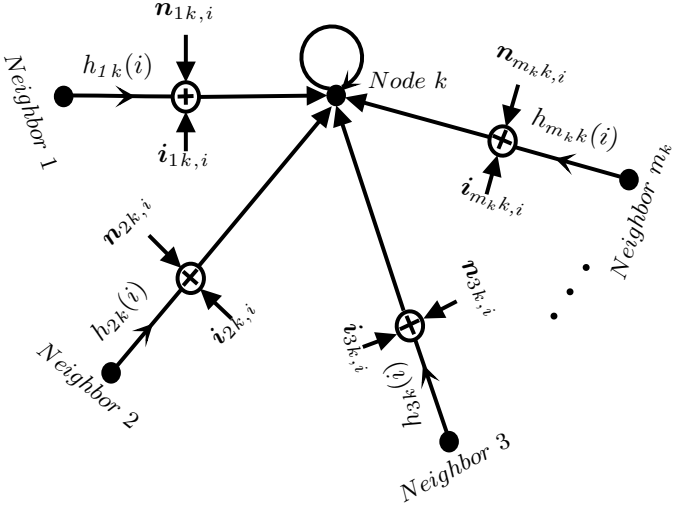
\begin{figure}[htb]
\centering

\tikzset{every picture/.style={line width=0.75pt}} 

\begin{tikzpicture}[x=0.75pt,y=0.75pt,yscale=-1,xscale=1]

\draw  [fill={rgb, 255:red, 0; green, 0; blue, 0 }  ,fill opacity=1 ] (174.99,242.67) .. controls (174.99,239.94) and (176.97,237.73) .. (179.42,237.73) .. controls (181.87,237.73) and (183.86,239.94) .. (183.86,242.67) .. controls (183.86,245.4) and (181.87,247.61) .. (179.42,247.61) .. controls (176.97,247.61) and (174.99,245.4) .. (174.99,242.67) -- cycle ;
\draw  [line width=1.5] [line join = round][line cap = round] (183.88,242.18) .. controls (200.96,242.18) and (216.24,242.67) .. (233.5,242.67) ;
\draw  [line width=1.5]  (234.41,242.92) .. controls (234.41,238.83) and (237.38,235.51) .. (241.06,235.51) .. controls (244.73,235.51) and (247.71,238.83) .. (247.71,242.92) .. controls (247.71,247.01) and (244.73,250.33) .. (241.06,250.33) .. controls (237.38,250.33) and (234.41,247.01) .. (234.41,242.92) -- cycle ;
\draw  [fill={rgb, 255:red, 0; green, 0; blue, 0 }  ,fill opacity=1 ] (207.8,238.97) -- (214.45,242.67) -- (207.8,246.37) -- (211.13,242.67) -- cycle ;
\draw [line width=1.5]    (247.71,242.92) -- (308.45,242.22) ;
\draw [shift={(312.45,242.18)}, rotate = 179.34] [fill={rgb, 255:red, 0; green, 0; blue, 0 }  ][line width=0.08]  [draw opacity=0] (10.45,-5.02) -- (0,0) -- (10.45,5.02) -- cycle    ;
\draw [line width=1.5]    (241.06,211.56) -- (241.06,231.51) ;
\draw [shift={(241.06,235.51)}, rotate = 270] [fill={rgb, 255:red, 0; green, 0; blue, 0 }  ][line width=0.08]  [draw opacity=0] (10.45,-5.02) -- (0,0) -- (10.45,5.02) -- cycle    ;
\draw [line width=1.5]    (241.06,270.33) -- (241.06,254.33) ;
\draw [shift={(241.06,250.33)}, rotate = 90] [fill={rgb, 255:red, 0; green, 0; blue, 0 }  ][line width=0.08]  [draw opacity=0] (10.45,-5.02) -- (0,0) -- (10.45,5.02) -- cycle    ;
\draw  [fill={rgb, 255:red, 0; green, 0; blue, 0 }  ,fill opacity=1 ] (312.45,242.18) .. controls (312.45,239.45) and (314.44,237.24) .. (316.89,237.24) .. controls (319.33,237.24) and (321.32,239.45) .. (321.32,242.18) .. controls (321.32,244.9) and (319.33,247.12) .. (316.89,247.12) .. controls (314.44,247.12) and (312.45,244.9) .. (312.45,242.18) -- cycle ;
\draw  [fill={rgb, 255:red, 0; green, 0; blue, 0 }  ,fill opacity=1 ] (198.93,382.18) .. controls (198.93,379.46) and (200.92,377.25) .. (203.37,377.25) .. controls (205.82,377.25) and (207.8,379.46) .. (207.8,382.18) .. controls (207.8,384.91) and (205.82,387.12) .. (203.37,387.12) .. controls (200.92,387.12) and (198.93,384.91) .. (198.93,382.18) -- cycle ;
\draw  [fill={rgb, 255:red, 0; green, 0; blue, 0 }  ,fill opacity=1 ] (369.4,416.92) .. controls (369.4,414.19) and (371.39,411.98) .. (373.84,411.98) .. controls (376.29,411.98) and (378.27,414.19) .. (378.27,416.92) .. controls (378.27,419.65) and (376.29,421.86) .. (373.84,421.86) .. controls (371.39,421.86) and (369.4,419.65) .. (369.4,416.92) -- cycle ;
\draw  [fill={rgb, 255:red, 0; green, 0; blue, 0 }  ,fill opacity=1 ] (460.11,284.65) .. controls (460.11,281.92) and (462.1,279.71) .. (464.55,279.71) .. controls (467,279.71) and (468.98,281.92) .. (468.98,284.65) .. controls (468.98,287.38) and (467,289.59) .. (464.55,289.59) .. controls (462.1,289.59) and (460.11,287.38) .. (460.11,284.65) -- cycle ;
\draw [line width=1.5]    (206.47,378.48) -- (238.84,339.96) ;
\draw  [line width=1.5]  (238.83,329.25) .. controls (241.43,326.35) and (245.64,326.35) .. (248.24,329.25) .. controls (250.84,332.14) and (250.84,336.83) .. (248.24,339.72) .. controls (245.64,342.62) and (241.43,342.62) .. (238.83,339.72) .. controls (236.23,336.83) and (236.23,332.14) .. (238.83,329.25) -- cycle ;
\draw  [line width=1.5]  (394.39,263.24) .. controls (395.34,259.29) and (398.99,256.95) .. (402.54,258.01) .. controls (406.08,259.07) and (408.19,263.13) .. (407.24,267.08) .. controls (406.29,271.03) and (402.64,273.37) .. (399.09,272.31) .. controls (395.55,271.25) and (393.44,267.19) .. (394.39,263.24) -- cycle ;
\draw  [line width=1.5]  (360.62,357.28) .. controls (361.22,361.29) and (358.77,365.08) .. (355.14,365.75) .. controls (351.52,366.42) and (348.1,363.71) .. (347.5,359.7) .. controls (346.9,355.69) and (349.36,351.9) .. (352.98,351.23) .. controls (356.6,350.56) and (360.03,353.27) .. (360.62,357.28) -- cycle ;
\draw [line width=1.5]    (247.71,328.11) -- (312.11,250.68) ;
\draw [shift={(314.67,247.61)}, rotate = 129.75] [fill={rgb, 255:red, 0; green, 0; blue, 0 }  ][line width=0.08]  [draw opacity=0] (10.45,-5.02) -- (0,0) -- (10.45,5.02) -- cycle    ;
\draw [line width=1.5]    (356.84,365.75) -- (372.56,411.98) ;
\draw [line width=1.5]    (350.89,350.76) -- (318.13,250.92) ;
\draw [shift={(316.89,247.12)}, rotate = 71.83] [fill={rgb, 255:red, 0; green, 0; blue, 0 }  ][line width=0.08]  [draw opacity=0] (10.45,-5.02) -- (0,0) -- (10.45,5.02) -- cycle    ;
\draw [line width=1.5]    (407.07,268.25) -- (460.11,283.17) ;
\draw [line width=1.5]    (394.39,263.24) -- (325.18,244.22) ;
\draw [shift={(321.32,243.16)}, rotate = 15.36] [fill={rgb, 255:red, 0; green, 0; blue, 0 }  ][line width=0.08]  [draw opacity=0] (10.45,-5.02) -- (0,0) -- (10.45,5.02) -- cycle    ;
\draw [line width=1.5]    (222.43,312.8) -- (236.01,326.41) ;
\draw [shift={(238.83,329.25)}, rotate = 225.09] [fill={rgb, 255:red, 0; green, 0; blue, 0 }  ][line width=0.08]  [draw opacity=0] (10.45,-5.02) -- (0,0) -- (10.45,5.02) -- cycle    ;
\draw [line width=1.5]    (264.12,354.78) -- (251.14,342.47) ;
\draw [shift={(248.24,339.72)}, rotate = 43.47] [fill={rgb, 255:red, 0; green, 0; blue, 0 }  ][line width=0.08]  [draw opacity=0] (10.45,-5.02) -- (0,0) -- (10.45,5.02) -- cycle    ;
\draw [line width=1.5]    (330.31,369.13) -- (343.81,364.1) ;
\draw [shift={(347.55,362.7)}, rotate = 159.54] [fill={rgb, 255:red, 0; green, 0; blue, 0 }  ][line width=0.08]  [draw opacity=0] (10.45,-5.02) -- (0,0) -- (10.45,5.02) -- cycle    ;
\draw [line width=1.5]    (377.66,346.06) -- (364.42,352.93) ;
\draw [shift={(360.87,354.78)}, rotate = 332.55] [fill={rgb, 255:red, 0; green, 0; blue, 0 }  ][line width=0.08]  [draw opacity=0] (10.45,-5.02) -- (0,0) -- (10.45,5.02) -- cycle    ;
\draw [line width=1.5]    (408.68,236.74) -- (403.65,254.16) ;
\draw [shift={(402.54,258)}, rotate = 286.11] [fill={rgb, 255:red, 0; green, 0; blue, 0 }  ][line width=0.08]  [draw opacity=0] (10.45,-5.02) -- (0,0) -- (10.45,5.02) -- cycle    ;
\draw [line width=1.5]    (392.71,292.06) -- (397.13,276.17) ;
\draw [shift={(398.21,272.31)}, rotate = 105.55] [fill={rgb, 255:red, 0; green, 0; blue, 0 }  ][line width=0.08]  [draw opacity=0] (10.45,-5.02) -- (0,0) -- (10.45,5.02) -- cycle    ;
\draw  [fill={rgb, 255:red, 0; green, 0; blue, 0 }  ,fill opacity=1 ] (216.62,360.63) -- (223.71,358.12) -- (221.26,365.94) -- (221.33,360.7) -- cycle ;
\draw  [fill={rgb, 255:red, 0; green, 0; blue, 0 }  ,fill opacity=1 ] (360.52,387.25) -- (361.06,379.03) -- (366.69,384.66) -- (362.33,382.5) -- cycle ;
\draw  [fill={rgb, 255:red, 0; green, 0; blue, 0 }  ,fill opacity=1 ] (433.88,279.79) -- (428.3,274.32) -- (435.57,272.63) -- (431.51,275.27) -- cycle ;
\draw [line width=1.5]    (240.97,238.61) -- (241.07,246.85) ;
\draw [line width=1.5]    (244.94,242.45) -- (236.9,242.75) ;
\draw [line width=1.5]    (239.76,338.1) -- (246.77,330) ;
\draw [line width=1.5]    (246.87,337.33) -- (239.66,330.99) ;
\draw [line width=1.5]    (349.01,360.19) -- (358.72,355.61) ;
\draw [line width=1.5]    (356.2,362.68) -- (352.34,353.42) ;
\draw [line width=1.5]    (395.88,262.44) -- (405.32,266.53) ;
\draw [line width=1.5]    (402.6,259.64) -- (398.72,269.05) ;
\draw  [draw opacity=0][line width=1.5]  (313.61,238.7) .. controls (310.89,238.51) and (308.18,237.57) .. (305.79,235.83) .. controls (298.9,230.81) and (297.28,221.01) .. (302.17,213.94) .. controls (307.06,206.87) and (316.61,205.21) .. (323.5,210.23) .. controls (330.4,215.25) and (332.02,225.05) .. (327.13,232.12) .. controls (325.25,234.83) and (322.69,236.75) .. (319.85,237.8) -- (314.65,223.03) -- cycle ; \draw  [line width=1.5]  (313.61,238.7) .. controls (310.89,238.51) and (308.18,237.57) .. (305.79,235.83) .. controls (298.9,230.81) and (297.28,221.01) .. (302.17,213.94) .. controls (307.06,206.87) and (316.61,205.21) .. (323.5,210.23) .. controls (330.4,215.25) and (332.02,225.05) .. (327.13,232.12) .. controls (325.25,234.83) and (322.69,236.75) .. (319.85,237.8) ;  
\draw  [fill={rgb, 255:red, 0; green, 0; blue, 0 }  ,fill opacity=1 ] (326.66,236.76) -- (320.15,238.92) -- (322.37,232.27) -- (322.33,236.72) -- cycle ;

\draw (331.43,228.46) node [anchor=north west][inner sep=0.75pt]  [font=\small,rotate=-0.58]  {$Node\ k$};
\draw (187.18,219.35) node [anchor=north west][inner sep=0.75pt]  [font=\normalsize]  {$\mathit{h_{1}{}_{k}}( i)$};
\draw (223.76,194.32) node [anchor=north west][inner sep=0.75pt]  [font=\normalsize]  {$\boldsymbol{n}_{1}{}_{k}{}_{,}{}_{i}$};
\draw (225.29,264.97) node [anchor=north west][inner sep=0.75pt]  [font=\normalsize]  {$\boldsymbol{i}_{1}{}_{k}{}_{,}{}_{i}$};
\draw (452.56,334.02) node [anchor=north west][inner sep=0.75pt]  [font=\small,rotate=-288.3]  {$Neighbor\ m_{k}$};
\draw (452.96,328.28) node [anchor=north west][inner sep=0.75pt]  [font=\Huge,rotate=-123.48]  {$\cdots $};
\draw (196.58,314.77) node [anchor=north west][inner sep=0.75pt]  [font=\normalsize,rotate=-311.24]  {$\boldsymbol{n}_{2}{}_{k}{}_{,}{}_{i}$};
\draw (249.34,362.4) node [anchor=north west][inner sep=0.75pt]  [font=\normalsize,rotate=-311.63]  {$\boldsymbol{i}_{2}{}_{k}{}_{,}{}_{i}$};
\draw (385.48,322.24) node [anchor=north west][inner sep=0.75pt]  [font=\normalsize,rotate=-61.48]  {$\boldsymbol{n}_{3}{}_{k}{}_{,}{}_{i}$};
\draw (329.72,353.17) node [anchor=north west][inner sep=0.75pt]  [font=\normalsize,rotate=-69.6]  {$\boldsymbol{i}_{3}{}_{k}{}_{,}{}_{i}$};
\draw (397.63,214.59) node [anchor=north west][inner sep=0.75pt]  [font=\normalsize,rotate=-18.87]  {$\boldsymbol{n}_{m_{k} k,i}$};
\draw (380.58,280.6) node [anchor=north west][inner sep=0.75pt]  [font=\normalsize,rotate=-16.71]  {$\boldsymbol{i}_{m_{k} k,i}$};
\draw (188.9,360.13) node [anchor=north west][inner sep=0.75pt]  [font=\normalsize,rotate=-311.29]  {$h_{2}{}_{k}( i)$};
\draw (349.8,413.55) node [anchor=north west][inner sep=0.75pt]  [font=\normalsize,rotate=-250.8]  {$h_{3}{}_{k}( i)$};
\draw (416.54,245.17) node [anchor=north west][inner sep=0.75pt]  [font=\normalsize,rotate=-15.45]  {$h_{m}{}_{_{k} k}( i)$};
\draw (158.08,217.41) node [anchor=north west][inner sep=0.75pt]  [font=\small,rotate=-71.12]  {$Neighbor\ 1$};
\draw (170.21,363.17) node [anchor=north west][inner sep=0.75pt]  [font=\small,rotate=-38.83]  {$Neighbor\ 2$};
\draw (344.85,440.06) node [anchor=north west][inner sep=0.75pt]  [font=\small,rotate=-331.89]  {$Neighbor\ 3$};

\end{tikzpicture}

\caption{The received signal at node $k$ considering fading and path loss, additive noise, and inter-node interference (INI).} 
\label{fig:M1}
\end{figure}

Consider a network of $K$ nodes which are distributed over a geographic region aiming at estimating an unknown vector $\boldsymbol{\omega}^{o}\in\mathbb{C}^{M}$. $\mathcal{N}_{k}$ denotes the set of neighbors of node $k$ (including $k$ itself), which are located within the transmission range ($r_{o}$) of node $k$. At each time instance $i \in \left\{1,2,..., T\right\}$, every node $k\in\left\{1,2,...,K\right\}$ collects scalar measurement $d_{k}(i)$ and a $1\times M$ regression vector $\boldsymbol{\mathit{u}}_{k,i}$ which are related to $\boldsymbol{\omega}^{o}$ via the following linear regression model: 
\begin{equation}
    \label{eq_1}
    {d}_{k}(i)=\boldsymbol{\mathit{u}}_{k,i}\boldsymbol{\omega} ^{o}+\mathit{v}_{k}(i),
\end{equation}
where $\mathit{v}_{k}(i)$ denotes the additive zero-mean white Gaussian measurement noise at node $k$, with variance $\sigma^{2}_{v,k}$. The regression vectors $\boldsymbol{\mathit{u}}_{k,i}$ are also zero-mean with covariance matrices ${\boldsymbol{R}_{u,k}}=\mathbb{E}\{\boldsymbol{\mathit{u}}^{*}_{k,i}\boldsymbol{\mathit{u}}_{k,i}\}$. According to the DLMS algorithm, the unknown vector $\boldsymbol{\omega}^{o}$ is distributively estimated by simultaneous exchange information among nodes over noisy wireless links, which are also under influence of fading and path loss. Apart from these perturbations, however, the data is also subject to inter-node interference (INI). As a result, the received signal $\boldsymbol{\psi}_{lk,i}\in\mathbb{C}^{M\times 1}$ at node $k$ from a neighboring node $l$ is modeled as:
\begin{equation}
\label{eq_2}
\boldsymbol{\psi}_{lk,i}=\beta_{lk}(i)\boldsymbol{\psi}_{l,i}+\boldsymbol{\mathit{i}}_{lk,i}+\boldsymbol{\mathit{n}}_{lk,i},
\end{equation}
 where $\boldsymbol{\psi}_{l,i}$ represents the transmitted signal from node $l$ at time instant $i$. Moreover, $\beta_{lk}(i)=h_{lk}(i)\sqrt{\frac{P_o}{r_{lk}^\alpha}}$ \cite{ref_12} represents analog transmission, where $h_{lk}(i)$ denotes the channel coefficient between nodes $l$ and $k$, $P_{o}$ is the power of transmitter signal, $r_{lk}$ is the distance between nodes $l$ and $k$, and $\alpha$ is the path loss exponent (see Figure \ref{fig:M1}). We assume that the links among nodes are spatially uncorrelated Rayleigh channels, which are i.i.d over time, and thus $h_{lk}(i)$ is zero-mean Gaussian with variance $\sigma^{2}_{h,lk}$, and consequently $\beta_{lk}(i)$ is also zero-mean Gaussian with variance $\sigma^{2}_{h,lk}\frac{P_{o}}{r_{lk}^{\alpha}}$. The vector $\boldsymbol{\mathit{i}}_{lk,i}\in\mathbb {C}^{M}$ is the INI between nodes $k$ and $l$ which is defined as the superimposition of signals transmitted by neighboring nodes of $k$ except node $l$, i.e., $\boldsymbol{\mathit{i}}_{lk,i}=\sum _{l^{\prime}\in\mathcal{N}_{k}\setminus\left \{k,l\right \}}\beta_{l^{\prime}k}(i)\boldsymbol{\psi}_{l^{\prime},i}$. In addition, $\boldsymbol{\mathit{n}}_{lk,i}$ represents the zero-mean additive white Gaussian noise (AWGN) vector with covariance matrix $\sigma^{2}_{n,lk}\boldsymbol{I}_{M}$.
At any time instant $i$, due to channel impairments, some links may fail. Therefore, to ensure reliable communication, only a subset of $\mathcal{N}_{k}$, denoted by $\mathcal{N}_{k,i}$, whose signal-to-interference-noise ratio (SINR) exceeds a pre-defined threshold value are allowed to send the signal to node $k$. Based on the defined model of the received signal in (\ref{eq_2}), SINR be written as:
 \begin{equation}
     \label{eq_3}
     \text{SINR}_{lk}(i)=\frac{\left|\beta_{lk}(i)\right|^{2}}{\sum_{l^{\prime}\in\mathcal{N}_{k,i}\setminus\left\{k,l\right\}}\left|\beta_{l^{\prime}k}(i)\right|^{2}+\sigma^{2}_{n,lk}}.
 \end{equation}

 \section{Diffusion LMS (DLMS) and Performance Analysis under INI}\label{sec_3}
 \subsection{Diffusion LMS (DLMS) Strategy}
 In this paper, we consider the adapt-then-combine (ATC) version of diffusion least mean square (DLMS). The extension to combine-then-adapt (CTA) algorithm is straightforward. According to the ATC version of DLMS we have \cite{ref_5}:
 \begin{equation}
     \label{eq_4}
     \begin{cases}
        \boldsymbol{\psi}_{k,i}=\boldsymbol{\omega}_{k,i-1}+\mu _{k}\boldsymbol{u}^{*}_{k,i}({d}_{k}(i)-\boldsymbol{u}_{k,i}\boldsymbol{\omega}_{k,i-1}),\quad \text{(Adaptation)} \\
        \boldsymbol{\omega}_{k,i}=\sum_{l\in \mathcal{N}_{k,i}}\mathit{a}_{lk}(i)\boldsymbol{\psi}_{lk,i}, \quad\text{(Combination)}
     \end{cases}
 \end{equation}
where $\mu_{k}$ is the step-size at node $k$, and $\left\{\mathit{a}_{lk}(i)\right\}$ are non-negative real coefficients corresponding to the entries of a left-stochastic combination matrix $\boldsymbol{A}_{i}$ such that
\begin{equation}
    \label{eq_5}
    \mathit{a}_{lk}(i)=0 \quad \text{if} \quad l\notin \mathcal{N}_{k,i} \quad \text{and} \sum_{l\in\mathit{N}_{k,i}}\mathit{a}_{lk}(i)=1,
\end{equation}
where, the index $``i"$ in $\boldsymbol{A}_{i}$ is for time-varying topology. 
To compensate for the fading and path-loss perturbations, we multiply the received signal $\boldsymbol{\psi}_{lk,i}$ by an equalization coefficient denoted as $\mathit{g}_{lk}(i)$ \cite{ref_11}. Therefore, from (\ref{eq_2}) the combination step of (\ref{eq_4}) is rewritten as 
\begin{equation}
    \label{eq_6}
           \boldsymbol{\omega}_{k,i}=\sum_{l\in\mathcal{N}_{k,i}}\mathit{q}_{lk}(i)\boldsymbol{\psi}_{l,i}+\boldsymbol{\mathit{i}}_{k,i}+\boldsymbol{\mathit{n}}_{k,i},
\end{equation}
where
\begin{equation}
    \label{eq_new_1}
    \mathit{q}_{lk}(i)\triangleq
    \begin{cases}
       \mathit{a}_{kk}(i)& l=k\\\mathit{a}_{lk}(i)\mathit{g}_{lk}(i)\beta_{lk}(i) &l\in\mathcal{N}_{k,i}\setminus\left\{k\right\},
    \end{cases}
\end{equation}
\subsection{Performance Analysis}
In this subsection, we analyze the steady-state behavior of the DLMS algorithm over the fading channel and in the presence of INI. The following assumptions are helpful for the analyses:

{\it Assumption 1:} All random processes $\boldsymbol{\mathit{n}}_{lk,i}$, $h_{lk}(i)$, $\mathit{v}_{k}(i)$, and  $\boldsymbol{\mathit{u}}_{k,i}$ are independent and identically distributed (i.i.d) over time and independent over space.

{\it Assumption 2:} The channel noise $\boldsymbol{\mathit{n}}_{lk,i}$, channel coefficients $h_{lk}(i)$, the measurement noise $\mathit{v}_{k}(i)$, and regression vectors  $\boldsymbol{\mathit{u}}_{k,i}$ are mutually independent.

{\it Assumption 3:} Based on the above assumptions, INI vector $\boldsymbol{\mathit{i}}_{lk,i}$ is independent of channel noise $\boldsymbol{\mathit{n}}_{lk,i}$, measurement noise $\mathit{v}_{k}(i)$, and regression vector $\boldsymbol{\mathit{u}}_{k,i}$, thus $\boldsymbol{\mathit{i}}_{lk,i}$ has zero-mean and covariance matrix $\boldsymbol{R}_{i,lk}\triangleq\left(\sum_{l^{\prime} \in \mathcal{N}_{k}\setminus\left\{l,k\right\}}\sigma^{2}_{h,l^{\prime}k}\frac{P_{o}}{r_{l^{\prime}k}^{\alpha}}\right)\boldsymbol{I}_{M}=\sigma^{2}_{i,lk}\boldsymbol{I}_{M}$, where $\sigma^{2}_{i,lk} = \sum_{l^{\prime} \in \mathcal{N}_{k}\setminus\left\{l,k\right\}}\sigma^{2}_{h,l^{\prime}k}\frac{P_{o}}{r_{l^{\prime}k}^{\alpha}}$.

Assuming that the vector $\boldsymbol{\omega}^{o}$ is invariant, then we define the error vectors $\boldsymbol{\tilde{\psi}}_{k,i}\triangleq\boldsymbol{\omega}^{o}-\boldsymbol{\psi}_{k,i}$ and  $\boldsymbol{\tilde{\omega}}_{k,i}\triangleq\boldsymbol{\omega}^{o}-\boldsymbol{\omega}_{k,i}$. Now, we subtract $\boldsymbol{\omega}^{o}$ from both sides of the adaptation step of (\ref{eq_4}) and (\ref{eq_6}) to obtain:
\begin{equation}
    \label{eq_7}
    \boldsymbol{\tilde{\psi}}_{k,i}=\left(\boldsymbol{I}_{M}-\mu_{k}\boldsymbol{\mathit{u}}_{k,i}^{*}\boldsymbol{\mathit{u}}_{k,i}\right)\boldsymbol{\tilde{\omega}}_{k,i-1}-\mu_{k}\boldsymbol{\mathit{u}}_{k,i}^{*}\mathit{v}_{k}(i),
\end{equation}
\begin{equation}
    \label{eq_8}
\boldsymbol{\tilde{\omega}}_{k,i}=\sum_{l\in\mathcal{N}_{k,i}}\mathit{q}_{lk}(i)\boldsymbol{\tilde{\psi}}_{l,i}+\sum_{l\in\mathcal{N}_{k,i}}\mathit{e}_{lk}(i)\boldsymbol{\omega}^{o}-\boldsymbol{\mathit{i}}_{k,i}-\boldsymbol{\mathit{n}}_{k,i},
\end{equation}
where 
\begin{equation}
    \label{eq_new2}
    \mathit{e}_{lk}(i)\triangleq\mathit{a}_{lk}(i)-\mathit{q}_{lk}(i).
\end{equation}
We also introduce the network global error vectors as:
\begin{equation}
  \boldsymbol{\tilde{\psi}}_{i}\triangleq\text{col}\left\{\boldsymbol{\tilde{\psi}}_{1,i},\boldsymbol{\tilde{\psi}}_{2,i},...,\boldsymbol{\tilde{\psi}}_{K,i}\right\}  
\end{equation}
\begin{equation}
  \boldsymbol{\tilde{\omega}}_{i}\triangleq \text{col}\left\{\boldsymbol{\tilde{\omega}}_{1,i},\boldsymbol{\tilde{\omega}}_{2,i},...,\boldsymbol{\tilde{\omega}}_{K,i}\right\} 
\end{equation}
 Moreover, we collect $\left\{\mathit{q}_{lk}(i)\right\}$ and $\left\{\mathit{e}_{lk}(i)\right\}$ into $\boldsymbol{Q}_{i}$ and $\boldsymbol{E}_{i}$, respectively. Then, the following variables are introduced: 
 \begin{equation}
    \boldsymbol{\mathcal{A}}_{i}\triangleq \boldsymbol{A}_{i}\otimes \boldsymbol{I}_{M} 
 \end{equation}
 \begin{equation}
    \boldsymbol{\mathcal{Q}}_{i}\triangleq \boldsymbol{Q}_{i}\otimes \boldsymbol{I}_{M} 
 \end{equation}
 \begin{equation}
 \label{eq_new3}
    \boldsymbol{\mathcal{E}}_{i}\triangleq \boldsymbol{E}_{i}\otimes \boldsymbol{I}_{M}\stackrel{(\ref{eq_new2})}{=}  \boldsymbol{\mathcal{A}}_{i}-\boldsymbol{\mathcal{Q}}_{i}
 \end{equation}
  \begin{equation}
    \boldsymbol{\mathcal{M}}\triangleq \text{diag}\left\{\mu_{1}\boldsymbol{I}_{M},\mu_{2}\boldsymbol{I}_{M},...,\mu_{K}\boldsymbol{I}_{M}\right\}
 \end{equation}
 \begin{equation}
     \boldsymbol{\mathcal{R}}_{u,i}\triangleq \text{diag}\left\{\boldsymbol{u}^{*}_{1,i}\boldsymbol{u}_{1,i},\boldsymbol{u}^{*}_{2,i}\boldsymbol{u}_{2,i},...,\boldsymbol{u}^{*}_{K,i}\boldsymbol{u}_{K,i}, \right\}
 \end{equation}
 \begin{equation}
     \boldsymbol{\mathit{z}}_{i}\triangleq\text{col}\left\{\boldsymbol{u}^{*}_{1,i}v_{1}(i),\boldsymbol{u}^{*}_{2,i}v_{2}(i),...,\boldsymbol{u}^{*}_{K,i}v_{K}(i) \right\}
 \end{equation}
 \begin{equation}
     \boldsymbol{\mathit{i}}_{i}\triangleq\text{col}\left\{\boldsymbol{\mathit{i}}_{1,i},\boldsymbol{\mathit{i}}_{2,i},...,\boldsymbol{\mathit{i}}_{K,i}\right\}
 \end{equation}
 \begin{equation}
     \boldsymbol{\mathit{n}}_{i}\triangleq\text{col}\left\{\boldsymbol{\mathit{n}}_{1,i},\boldsymbol{\mathit{n}}_{2,i},...,\boldsymbol{\mathit{n}}_{K,i}\right\}
 \end{equation}
 \begin{equation}
     \boldsymbol{\mathcal{\omega}}_{c}^{o}\triangleq\mathbf{1}_{K}\otimes\boldsymbol{\omega}^{o}
 \end{equation}
  From these variables along with (\ref{eq_7}) and (\ref{eq_8}) the network error vector is obtained as
\begin{equation}
    \label{eq_9}
\boldsymbol{\tilde{\omega}}_{i}=\boldsymbol{\mathcal{B}}_{i}\boldsymbol{\tilde{\omega}}_{i-1}-\boldsymbol{\mathcal{Q}}_{i}^{T}\boldsymbol{\mathcal{M}}\boldsymbol{\mathit{z}}_{i}+\boldsymbol{\mathcal{E}}_{i}\boldsymbol{\mathcal{\omega}}_{c}^{o}-\boldsymbol{\mathit{i}}_{i}-\boldsymbol{\mathit{n}}_{i},
\end{equation}
where $\boldsymbol{\mathcal{B}}_{i}\triangleq\boldsymbol{\mathcal{Q}}_{i}^{T}\left(\boldsymbol{I}_{MK}-\boldsymbol{\mathcal{M}}\boldsymbol{\mathcal{R}}_{u,i}\right)$. 

\textit{Mean Convergence:} Taking expectation from both sides of (\ref{eq_9}), with the assumption $\mathbb{E}\{{\boldsymbol{\mathit{i}}_{i}}\}=\mathbb{E}\{{\boldsymbol{\mathit{n}}_{i}}\}=\mathbb{E}\{{\boldsymbol{\mathit{z}}_{i}}\}=0$, the following recursion for network mean error vector is obtained
\begin{equation}
    \label{eq_10}
\mathbb{E}\{\boldsymbol{\tilde{\omega}}_{i}\}=\boldsymbol{\mathcal{B}}\mathbb{E}\{\boldsymbol{\tilde{\omega}}_{i-1}\}+\boldsymbol{\mathcal{E}}\boldsymbol{\mathcal{\omega}}_{c}^{o},
\end{equation}
where $\boldsymbol{\mathcal{B}}=\mathbb{E}\{\boldsymbol{\mathcal{B}}_{i}\}$ and $\boldsymbol{\mathcal{E}}=\mathbb{E}\{\boldsymbol{\mathcal{E}}_{i}\}$. Based on (\ref{eq_10}), if $\boldsymbol{\mathcal{B}}$ is stable, the network mean error vector will converge to
\begin{equation}
    \label{eq_11}
\lim_{i\rightarrow\infty}\mathbb{E}\{\boldsymbol{\tilde{\omega}}_{i}\}=\left(\boldsymbol{I}_{MK}-\boldsymbol{\mathcal{B}}\right)^{-1}\boldsymbol{\mathcal{E}}\boldsymbol{\mathcal{\omega}}_{c}^{o}.
\end{equation}
As (\ref{eq_11}) reveals, when the fading channel exists the algorithm is not asymptotically unbiased unless the equalizer coefficient $\mathit{g}_{lk}(i)$ is applied to the channel coefficient. For instance, if $\mathit{g}_{lk}(i)$ is zero-forcing (ZF) equalizer, i.e., $\mathit{g}_{lk}(i)=\frac{\beta_{lk}^{*}(i)}{\left|\beta_{lk}(i)\right|^{2}}$, then 
since $\mathit{q}_{lk}(i)=\mathit{a}_{lk}(i)\mathit{g}_{lk}(i)\beta_{lk}(i)$ we have $\mathit{q}_{lk}(i)=\mathit{a}_{lk}(i)$ and $\boldsymbol{\mathcal{Q}}_{i}^{T}=\boldsymbol{\mathcal{A}}_{i}^{T}$. Hence, according to (\ref{eq_new3}) we can deduce that $\boldsymbol{\mathcal{E}}=\boldsymbol{0}_{MK}$.
As a result, (\ref{eq_11}) converges to zero, and thus the algorithm is asymptotically unbiased. Likewise, for the ideal channel, i.e., $\beta_{lk}(i)=1$ for any $l$ and $k$, $\boldsymbol{\mathcal{E}}=\boldsymbol{0}_{MK}$, which gives the same result as the condition in which ZF equalizer is used. We now derive the condition, under which $\boldsymbol{\mathcal{B}}$ is stable, i.e., $\rho\left(\boldsymbol{\mathcal{B}}\right)<1$, where $\rho\left(\boldsymbol{\mathcal{B}}\right)$ represents the spectral radius of $\boldsymbol{\mathcal{B}}$. To this end, we use the block maximum norm properties as \cite{ref_27}:
 \begin{equation}
     \label{eq_12}
     \begin{split}
      \rho\left(\mathcal{B}\right)\leq\left\|\mathcal{B}\right\|_{b,\infty}=&\left\|\mathcal{Q}^{T}\left(I_{MN}-\mathcal{M}\mathcal{R}_{u}\right)\right\|_{b,\infty}\\&\leq\left\|\mathcal{Q}^{T}\right\|_{b,\infty}\left\|I_{MN}-\mathcal{M}\mathcal{R}_{u}\right\|_{b,\infty}
     \end{split}
 \end{equation}
where $\boldsymbol{\mathcal{Q}}=\mathbb{E}\{\boldsymbol{\mathcal{Q}}_{i}\}$ and $\boldsymbol{\mathcal{R}}_{u}=\mathbb{E}\{\boldsymbol{\mathcal{R}}_{u,i}\}$. Therefore, if $\left\|\boldsymbol{I}_{MN}-\boldsymbol{\mathcal{M}}\boldsymbol{\mathcal{R}}_{u}\right\|_{b,\infty}<\frac{1}{\left\|\boldsymbol{\mathcal{Q}}^{T}\right\|_{b,\infty}}$, then $\rho\left(\boldsymbol{\mathcal{B}}\right)<1$. Since $\boldsymbol{I}_{MN}-\boldsymbol{\mathcal{M}}\boldsymbol{\mathcal{R}}_{u}$ is block diagonal Hermitian, $\delta=\rho(\boldsymbol{I}_{MN}-\boldsymbol{\mathcal{M}}\boldsymbol{\mathcal{R}}_{u})=\left\|\boldsymbol{I}_{MN}-\boldsymbol{\mathcal{M}}\boldsymbol{\mathcal{R}}_{u}\right\|_{b,\infty}$. Hence, if $\delta<\frac{1}{\left\|\boldsymbol{\mathcal{Q}}^{T}\right\|_{b,\infty}}$, then $\rho\left(\boldsymbol{\mathcal{B}}\right)<1$.
To satisfy the condition $\delta<\frac{1}{\left\|\boldsymbol{\mathcal{Q}}^{T}\right\|_{b,\infty}}$, we should have $\lambda_{\text{max}}\left(\boldsymbol{R}_{u,k}\right)<\frac{1}{\left\|\mathcal{Q}^{T}\right\|_{b,\infty}}$, where $\boldsymbol{R}_{u,k}=\mathbb{E}\{\boldsymbol{u}^{*}_{k,i}\boldsymbol{u}_{k,i}\}$. As a result, the step-size $\mu_{k}$ is chosen according to the following condition:
\begin{equation}
   \label{eq_13}
    \frac{1-\frac{1}{\left\|\boldsymbol{\mathcal{Q}}^{T}\right\|_{b,\infty}}}{\lambda_{\text{max}}\left(\boldsymbol{R}_{u,k}\right)}<\mu_{k}<\frac{1+\frac{1}{\left\|\boldsymbol{\mathcal{Q}}^{T}\right\|_{b,\infty}}}{\lambda_{\text{max}}\left(\boldsymbol{R}_{u,k}\right)}.
\end{equation}
For ideal channel or in the case of using ZF equalizer, because $\boldsymbol{\mathcal{Q}}^{T}=\boldsymbol{\mathcal{A}}^{T}$ and matrix $\boldsymbol{\mathcal{A}}$ is left-stochastic, i.e., $\|\boldsymbol{\mathcal{A}}^{T}\|_{b,\infty}=1$,  (\ref{eq_13}) reduces to $0<\mu_{k}<\frac{2}{\lambda_{\text{max}}\left(\boldsymbol{R}_{u,k}\right)}$, which is the mean stability condition for diffusion LMS over ideal communication channels \cite{ref_5}, \cite{ref_27}.

\textit{Mean-square Performance}: To study mean-square performance the variance relation of the network error vector is obtained. To that end, we rearrange (\ref{eq_9}) and we take the expected value of the weighted vector norm of both sides of (\ref{eq_9}) given Assumption 1, Assumption 2, and Assumption 3. We obtain:
\begin{equation}
    \label{eq_14}
    \begin{split}  \mathbb{E}\{\left\|\boldsymbol{\tilde{\omega}}_{i}\right\|^{2}_{\boldsymbol{\Sigma}}\}&=\mathbb{E}\{\left\|\boldsymbol{\tilde{\omega}}_{i-1}\right\|^{2}_{\boldsymbol{\Sigma}^{\prime}}\}+\text{Tr}(\mathbb{E}\{\boldsymbol{\mathcal{Q}}_{i}^{T}\boldsymbol{\mathcal{M}}\boldsymbol{\mathit{z}}_{i}\boldsymbol{\mathit{z}}^{*}_{i}\boldsymbol{\mathcal{M}}\boldsymbol{\mathcal{Q}}_{i}\boldsymbol{\Sigma}\})\\
    &\quad{}+\text{Tr}(\mathbb{E}\{\boldsymbol{\mathcal{E}}_{i}\boldsymbol{\mathcal{\omega}}_{c}^{o}\boldsymbol{\omega}_{c}^{o*}\boldsymbol{\mathcal{E}}_{i}^{T}\boldsymbol{\Sigma}\})\\
    &\quad{}+2\text{Re}\{\text{Tr}(\mathbb{E}\{\boldsymbol{\mathcal{B}}_{i}\boldsymbol{\tilde{\omega}}_{i-1}\boldsymbol{\mathcal{\omega}}_{c}^{o}\boldsymbol{\mathcal{E}}^{T}_{i}\boldsymbol{\Sigma}\})\}\\
    &\quad{}+\text{Tr}\left(\mathbb{E}\{\boldsymbol{\mathit{i}}_{i}\boldsymbol{\mathit{i}}_{i}^{*}\boldsymbol{\Sigma}\}\right)+\text{Tr}\left(\mathbb{E}\{\boldsymbol{\mathit{n}}_{i}\boldsymbol{\mathit{n}}_{i}^{*}\boldsymbol{\Sigma}\}\right),
    \end{split}
\end{equation}
where $\boldsymbol{\Sigma}$ can be any Hermitian positive-definite matrix and $\Sigma^{\prime}=\mathbb{E}\{\boldsymbol{\mathcal{B}}^{*}_{i}\boldsymbol{\Sigma}\boldsymbol{\mathcal{B}}_{i}\}$.
Let $\boldsymbol{\sigma}\triangleq\operatorname{vec}\left(\boldsymbol{\Sigma}\right)$. We also use the notation $\mathbb{E}\{\left\|\boldsymbol{\tilde{\omega}}_{i}\right\|_{\boldsymbol{\sigma}}^{2}\}$ to denote $\mathbb{E}\{\left\|\boldsymbol{\tilde{\omega}}_{i}\right\|_{\boldsymbol{\Sigma}}^{2}\}$. Using some algebra such as $\text{vec}\left(\boldsymbol{U}\boldsymbol{\Sigma} \boldsymbol{V}\right)=\left(\boldsymbol{V}^{T}\otimes \boldsymbol{U}\right)\text{vec}\left(\boldsymbol{\Sigma}\right)$ and $\text{Tr}\left(\boldsymbol{\Sigma} \boldsymbol{X}\right)=\text{vec}\left(\boldsymbol{X}^{T}\right)^{T}\boldsymbol{\sigma}$, we obtain the following recursion for the network error variance:
 \begin{equation}
     \label{eq_15}
     \begin{split}
     \mathbb{E}\left\|\tilde{\boldsymbol{\omega}}_{i}\right\|^{2}_{\sigma}&= \mathbb{E}\left\|\tilde{\boldsymbol{\omega}}_{i-1}\right\|^{2}_{\mathcal{F}\sigma}+\gamma^{T}\sigma\\&=\mathbb{E}\left\|\boldsymbol{\tilde{\omega}}_{-1}\right\|^{2}_{\mathcal{F}^{i+1}\sigma}+\gamma^{T}\sum_{j=0}^{i}\mathcal{F}^{j}\sigma
     \end{split}
 \end{equation}
where $\boldsymbol{\mathcal{F}}=\boldsymbol{\mathcal{B}}^{T}\otimes\boldsymbol{\mathcal{B}}^{*}$ and $\boldsymbol{\gamma}$ is determined as:
\begin{equation}
\label{eq_16}
\begin{split}
     \boldsymbol{\gamma}&=2\text{Re}\left\{\text{vec}\left(\boldsymbol{\mathcal{B}}\mathbb{E}\{\boldsymbol{\tilde{\omega}}_{i-1}\boldsymbol{\mathcal{\omega}}^{o*}\}\boldsymbol{\mathcal{E}}^{T}\right)\right\}+\text{vec}\left(\boldsymbol{\mathcal{E}}\boldsymbol{\mathcal{\omega}}^{o*}\boldsymbol{\mathcal{\omega}}^{o}\boldsymbol{\mathcal{E}}\right)\\
&\quad{}+\text{vec}\left(\boldsymbol{\mathcal{Q}}^{T}\boldsymbol{\mathcal{M}}\boldsymbol{\mathcal{Z}}^{T}\boldsymbol{\mathcal{M}}\boldsymbol{\mathcal{Q}}\right)+\text{vec}\left(\boldsymbol{\mathcal{R}}_{int}^{T}\right)+\text{vec}\left(\boldsymbol{\mathcal{R}}_{n}^{T}\right),
\end{split}
\end{equation}
where $\boldsymbol{\mathcal{R}}_{int}=\text{diag}\left\{\boldsymbol{R}_{int,1},\boldsymbol{R}_{int,2},...,\boldsymbol{R}_{int,K}\right\}$, 
$\boldsymbol{\mathcal{Z}}=\mathbb{E}\{\boldsymbol{\mathit{z}}_{i}\boldsymbol{\mathit{z}}_{i}^{*}\}=\text{diag}\left\{\boldsymbol{R}_{u,1}\sigma^{2}_{v,1},\boldsymbol{R}_{u,2}\sigma^{2}_{v,2},...,\boldsymbol{R}_{u,K}\sigma^{2}_{v,K}\right\}$, and $\boldsymbol{\mathcal{R}}_{n}=\text{diag}\left\{\boldsymbol{R}_{n,1},\boldsymbol{R}_{n,2},...,\boldsymbol{R}_{n,K}\right\}$. In addition, $\boldsymbol{R}_{int,k}=\mathbb{E}\{\boldsymbol{\mathit{i}}_{k,i}\boldsymbol{\mathit{i}}_{k,i}^{*}\}=\sum_{l\in \mathcal{N}_{k}\setminus\left\{k\right\}}\mathbb{E}\{\mathit{a}_{lk}^{2}(i)\left|\mathit{g}_{lk}(i)\right|^{2}\}\boldsymbol{R}_{i,lk}$, and $\boldsymbol{R}_{n,k}=\mathbb{E}\{\boldsymbol{\mathit{n}}_{k,i}\boldsymbol{\mathit{n}}_{k,i}^{*}\}=\sum_{l\in \mathcal{N}_{k}\setminus\left\{k\right\}}\mathbb{E}\{\mathit{a}_{lk}^{2}(i)\left|\mathit{g}_{lk}(i)\right|^{2}\}\boldsymbol{R}_{n,lk}$, where $\boldsymbol{R}_{n,lk}=\sigma^{2}_{n,lk}\boldsymbol{I}_{M}$. 

The instantaneous mean square deviation (MSD) at node $k$, denoted by $\eta_{k}(i)$, is defined as $\eta_{k}(i)\triangleq\mathbb{E}\{\left\|\boldsymbol{\tilde{\omega}}_{i}\right\|^{2}\}$. Since we are free to choose $\boldsymbol{\sigma}$, $\eta_{k}(i)$ can be computed from (\ref{eq_15}) by selecting $\boldsymbol{\sigma}_{\text{msd}_{k}}=\text{vec}\left(\text{diag}\left(\boldsymbol{e}_{k}\right)\otimes \boldsymbol{I}_{M}\right)$, where $\boldsymbol{e}_{k}$ is a column vector with a unit element at position $k$ and zero elsewhere \cite{ref_5}. Therefore, using (\ref{eq_15}) and assuming $\boldsymbol{\omega}_{k,-1}=\boldsymbol{0}_{M}$ the instantaneous MSD at node $k$ is obtained as
\begin{equation}
    \label{eq_17}
    \eta_{k}(i)=\eta_{k}(i-1)-\left\|\boldsymbol{\mathcal{\omega}}^{o}\right\|^{2}_{\boldsymbol{\mathcal{F}}^{i}\left(\boldsymbol{I}_{MN}-\boldsymbol{\mathcal{F}}\right)\boldsymbol{\sigma}_{\text{msd}_{k}}}+\boldsymbol{\gamma}^{T}\boldsymbol{\mathcal{F}}^{i}\boldsymbol{\sigma}_{\text{msd}_{k}}.
\end{equation}
Eventually, from (\ref{eq_17}), the instantaneous network MSD will be derived as: $\eta(i)=\frac{1}{K}\sum_{k=1}^{K}\eta_{k}(i)$.  
\section{Optimized Combination Weights}
In this section, we derive an optimal combination rule by solving an optimization problem that is built on an upper bound of steady-state network MSD. To this end, first, similar to \cite{ref_12}, the entries of the time-varying matrix $\boldsymbol{A}_{i}$ are considered as $\mathit{a}_{lk}(i)=\zeta_{lk}\Gamma_{lk}(i)$, where $\Gamma_{lk}(i)$ is a random function with two possible values 0 and 1 for $l\notin\mathcal{N}_{k,i}$ and $l\in\mathcal{N}_{k,i}$, respectively, where $\sum_{l\in\mathcal{N}_{k,i}\setminus\left\{k\right\}}\zeta_{lk}<1$. The random function $\Gamma_{lk}(i)$ has binomial distribution with the probability of success $p_{lk}=\mathrm{Pr}\left(\text{SINR}_{lk}(i)\geq\text{SINR}_{th}\right)$ for successful transmission, where $\text{SINR}_{th}$ is the predefined threshold for the signal-to-interference-noise ratio. Second, we evaluate (\ref{eq_15}) when $i\rightarrow\infty$. At steady-state, because of the stability of matrix $\boldsymbol{\mathcal{B}}$, the matrix $\boldsymbol{\mathcal{F}}=\boldsymbol{\mathcal{B}}^{T}\otimes\boldsymbol{\mathcal{B}}^{*}$ will also be stable, so the first term of (\ref{eq_15}) approaches zero. However, the expression for $\boldsymbol{\gamma}$ includes unknown parameter $\boldsymbol{\omega}^{o}$, so we consider the assumption of using ZF equalizer which leads to $\boldsymbol{\mathcal{E}}=0$. Hence, substituting $\mathcal{F}$ into (\ref{eq_15}), setting $\boldsymbol{\sigma}=\frac{1}{K}\boldsymbol{I}_{MK}$, and using assumptions we obtain:
\begin{equation}
   \label{eq_18}
       \eta\approx\frac{1}{K}\sum_{j=0}^{\infty}\Tr\left[\boldsymbol{\mathcal{B}}^{j}\left(\boldsymbol{\mathcal{A}}^{T}\boldsymbol{\mathcal{M}}\boldsymbol{\mathcal{Z}}^{T}\boldsymbol{\mathcal{M}}\boldsymbol{\mathcal{A}}+\boldsymbol{\mathcal{R}}_{int}+\boldsymbol{\mathcal{R}}_{n}\right)\boldsymbol{\mathcal{B}}^{*j}\right].
\end{equation}
Using nuclear norm properties results in the following upper bound for $\eta$ \cite{ref_23}:
\begin{equation}
    \label{eq_19}
    \eta\leq\frac{c^{2}}{K}\frac{\Tr\left(\left(\boldsymbol{\mathcal{A}}^{T}\boldsymbol{\mathcal{M}}\boldsymbol{\mathcal{Z}}^{T}\boldsymbol{\mathcal{M}}\boldsymbol{\mathcal{A}}\right)+\boldsymbol{\mathcal{R}}_{int}+\boldsymbol{\mathcal{R}}_{n}\right)}{1-\left\|\boldsymbol{I}_{MN}-\boldsymbol{\mathcal{M}}\boldsymbol{\mathcal{R}}_{u}\right\|_{b,\infty}^{2j}},
\end{equation}
where $c$ is a positive scalar so that $\left \| \boldsymbol{X} \right \|_{*}\leq\left \| \boldsymbol{X} \right \|_{b,\infty}$ and $\left \| \boldsymbol{X} \right \|_{*}$ denotes the nuclear norm, which is defined as the sum of the singular values of $\boldsymbol{X}$. Therefore, $\|\boldsymbol{X}\|_*=\left\|\boldsymbol{X}^*\right\|_*$ for any $\boldsymbol{X}$ and $\|\boldsymbol{X}\|_*=\operatorname{Tr}(\boldsymbol{X})$ when $\boldsymbol{X}$ is Hermitian and positive semi-definite \cite{ref_23}.
This upper bound will be minimized if its numerator is minimized. Hence, we obtain the following element-wise problem for each node $k$:
\begin{equation}
\label{eq_20}
\begin{aligned}
\min_{\zeta_{lk}} & \sum_{l\in N_{k}}\zeta_{lk}^{2}p_{lk}\left[\mu_{l}^{2}\sigma^{2}_{v,l}\Tr\left(\boldsymbol{R}_{u,l}\right)+
\left|\mathit{g}_{lk}\right|^{2}M\left(\sigma^{2}_{i,lk}+\sigma^{2}_{n,lk}\right)\right],\\
\textrm{s.t.} \quad & \zeta_{lk}\geq0 \quad , \sum_{l\in \mathcal{N}_{k}}p_{lk}\zeta_{lk}=1 \quad , \zeta_{lk}=0 \quad \text{if} \quad l\notin \mathcal{N}_{k},
\end{aligned}
\end{equation}
where $\left|g_{lk}\right|^{2}=\mathbb{E}\{\left|\mathit{g}_{lk}(i)\right|^{2}\mid\text{SINR}_{lk}(i)\geq\text{SINR}_{th}\}$, which can be computed numerically over repetitious independent experiments, and  $\zeta^{2}_{lk}p_{lk}=\mathbb{E}\{\mathit{a}_{lk}(i)\}$. The minimization problem (\ref{eq_20}) is convex for the following reasons. First, since $\zeta_{lk}$ is a positive variable, $\zeta_{lk}^{2}$ is convex. Moreover, the linear constraints form a convex region \cite{ref_28}. As a result, using Lagrange dual function and applying Karush-Kuhn-Tucker (KKT) \cite{ref_28}, and $\mathbb{E}\left[\Gamma_{lk}(i)\right]=p_{lk}$, we derive a solution for the entries $\boldsymbol{A}_{i}$, which are the instantaneous combination weights as:
\begin{equation}
    \label{eq_21}
    \mathit{a}_{lk}(i)=
     \begin{cases}
         \frac{\alpha_{lk}^{-2}}{\sum_{m\in\mathcal{N}_{k,i}}\alpha^{-2}_{mk}} & l\in\mathcal{N}_{k,i}, \\
       0 & \text{otherwise},
    \end{cases}
\end{equation}
where
\begin{equation}
    \label{eq_22}
    \alpha_{lk}^{2}=
    \begin{cases}
    \begin{split}
        &\mu_{l}^{2}\sigma^{2}_{v,l}\Tr\left(\boldsymbol{R}_{u,l}\right)\\
        &\quad{}+\left|\mathit{g}_{lk}\right|^{2}M\left(\sigma^{2}_{i,lk}+\sigma^{2}_{n,lk}\right)
    \end{split}
        &  l\in\mathcal{N}_{k,i}\setminus\left\{k\right\},\\\\ \mu_{k}^{2}\sigma^{2}_{v,k}\Tr\left(\boldsymbol{R}_{u,k}\right) & l=k.
    \end{cases}
\end{equation}
As (\ref{eq_22}) reveals, to compute combination weights $\left\{\mathit{a}_{lk}(i)\right\}$ we need to have second-order moments $\left\{\sigma^{2}_{v,l},\Tr\left(\boldsymbol{R}_{u,l}\right),\left|\mathit{g}_{lk}\right|^{2},\sigma^{2}_{i,lk},\sigma^{2}_{n,lk}\right\}$, which are not available in practice. To overcome this challenge, we estimate $\alpha_{lk}^{2}$ using available information at every node. To that end, under Assumption 1-3, using ZF forcing equalizer, and utilizing (\ref{eq_1}), (\ref{eq_2}), and (\ref{eq_4}) for $l\in\mathcal{N}_{k,i}\setminus\left\{k\right\}$ we can write:
 \begin{equation}
     \label{eq_23}
     \begin{split}
         \alpha_{lk}^{2}&=\mathbb{E}\left\|\mathit{g}_{lk}(i)\boldsymbol{\psi}_{lk}(i)-\boldsymbol{\omega}_{l,i-1}\right\|^{2}\approx\mu_{l}^{2}\sigma^{2}_{v,l}\Tr\left(R_{u,l}\right)\\
         &\hspace{2cm}+M\mathbb{E}\left|\mathit{g}_{lk}(i)\right|^{2}\left(\sigma^{2}_{i,lk}+\sigma^{2}_{n,lk}\right)
     \end{split}
 \end{equation}
For $l=k$, as $\sigma^{2}_{i,lk}$ and $\sigma^{2}_{n,lk}$ are zero, we have:
\begin{equation}
    \label{eq_24}
    \alpha_{kk}^{2}=\mathbb{E}\left\|\boldsymbol{\psi}_{k,i}-\boldsymbol{\omega}_{k,i-1}\right\|^{2}\approx\mu_{k}^{2}\sigma^{2}_{k,v}\Tr\left(\boldsymbol{R}_{u,k}\right).
\end{equation}
Hence, using ZF equalizer, the proposed DLMS algorithm, under INI, converges in mean and mean-square sense, i.e., all estimates $\left\{\boldsymbol{\omega}_{k,i}\right\}$ converge to $\boldsymbol{\omega}^{o}$ as $i\rightarrow\infty$.

To estimate the adaptive combination coefficient $\alpha_{lk}^{2}(i)$ by using instantaneous realizations of  $\left\|\mathit{g}_{lk}(i)\boldsymbol{\psi_{lk,i}}-\boldsymbol{\omega}_{l,i-1}\right\|^{2}$, we replace  $\boldsymbol{\omega}_{l,i-1}$ (which is not available at node $k$) with $\boldsymbol{\omega}_{k,i-1}$, i.e, $\left\|\mathit{g}_{lk}(i)\boldsymbol{\psi_{lk,i}}-\boldsymbol{\omega}_{k,i-1}\right\|^{2}$. Similarly, instantaneous realizations of $\left\|\boldsymbol{\psi}_{k,i}-\boldsymbol{\omega}_{k,i-1}\right\|^2$ can be used to estimate $\alpha_{kk}^{2}(i)$. Moreover, as the network has a time-varying topology, we store $\alpha_{lk}^{2}(i-1)$ to recall them in next iterations. Ultimately, in light of these explanations, we propose an adaptive combination rule for such a network as:
\begin{equation}
    \label{eq_25}
    \mathit{a}_{lk}(i)=
    \begin{cases}
       \frac{\hat{\alpha}_{lk}^{-2}(i)}{\sum_{m\in\mathcal{N}_{k,i}}\hat{\alpha}^{-2}_{mk}(i)}& l\in\mathcal{N}_{k,i},\\0 &\text{otherwise},
    \end{cases}
\end{equation}
where $\hat{\alpha}^{2}_{lk}(i)$ is an estimation of $\alpha_{lk}^{2}(i)$ which is computed as:
\begin{equation}
    \label{eq_26}
    \hat{\alpha}^{2}_{lk}(i)=
    \begin{cases}
    \begin{split}
    &(1-\tau)\hat{\alpha}_{lk}^{2}(i-1)\\
    &\quad{}+\tau\left\|\mathit{g}_{lk}(i)\boldsymbol{\psi}_{lk}(i)-\boldsymbol{\omega}_{k,i-1}\right\|^2
    \end{split}
      & l\in\mathcal{N}_{k,i},\\\\
      \hat{\alpha}_{lk}^{2}(i-1) & l\notin\mathcal{N}_{k,i},
    \end{cases}
\end{equation}
where $0<\tau<1$ is the learning factor. Although the adaptive combination rule in (\ref{eq_25}) and (\ref{eq_26}) looks similar to (32) in \cite{ref_12}, however, in this paper, we examine a different scenario from \cite{ref_12}, in which the possible INI among nodes has been taken into account. Hence, for calculating $\hat{\alpha}^{2}_{lk}(i)$, which is the estimation of $\alpha_{lk}^{2}(i)$, we use (\ref{eq_23}), which is different from (32) in \cite{ref_12} because it requires the variance of INI, i.e., $\sigma^{2}_{i,lk}$.
\section{Simulation Results}\label{sec_6}
\begin{figure}[t]
     \centering
    \includegraphics[width=0.6\linewidth]{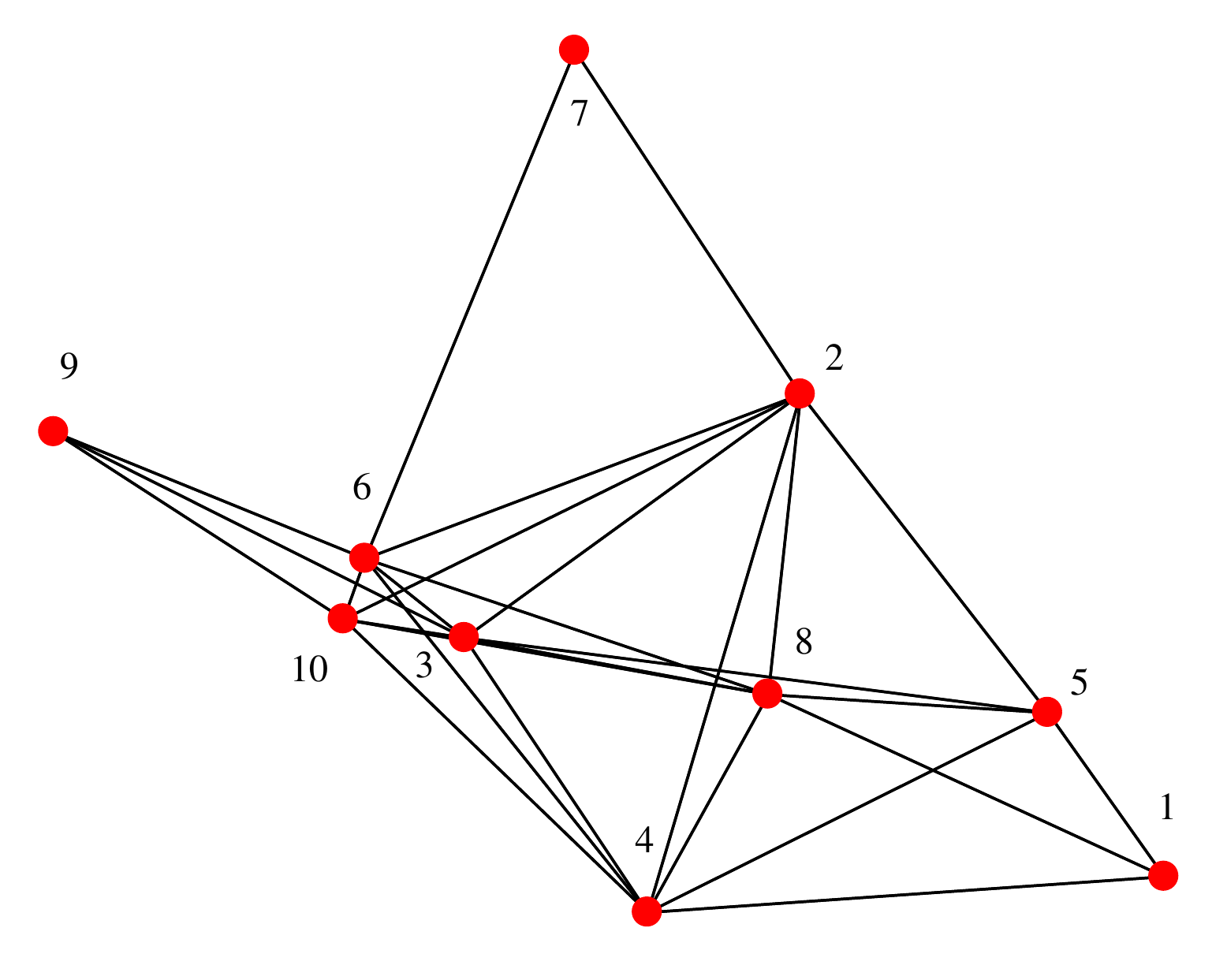}
    \caption{\footnotesize Topology of the wireless sensor network.}
     \label{figtopo}
\end{figure}

This section presents the simulation results to show the performance of DLMS algorithm over the wireless networks with fading channels, additive noise, and INI for different combination schemes. We consider a network with $K=10$ nodes. The topology of the network is shown in Figure \ref{figtopo}. The unknown parameter is assumed to be $\boldsymbol{\omega}^{o}=\left[1+j,-0.5-0.5j\right]^{T}$. We set the initial estimation vectors to $\boldsymbol{\omega}_{k,-1}=\mathbf{0}_{M}$. In addition, transmitter power $P_{o}=1$, transmission range $r_{o}=0.5$, path-loss exponent $\alpha=2.5$, and $\text{SINR}_{th}=-10dB$. The step size has been set to $\mu_{k}=0.01$ for all nodes. We used zero-mean complex circular Gaussian distribution to generate signals $\boldsymbol{\mathit{v}}_{k}(i)$, $\boldsymbol{\mathit{n}}_{lk,i}$ and $\boldsymbol{\mathit{u}}_{k,i}$. They have (co)variance $\sigma^{2}_{k,v}$, $R_{n,lk}=\sigma^{2}_{n,lk}I_{M}$, and $R_{u,k}$, respectively. We further generate the channel coefficients $\left\{\boldsymbol{\mathit{h}}_{lk}(i)\right\}$ based on zero-mean complex circular Gaussian distribution with variance $\sigma^{2}_{h,lk}=1$. We conduct the experiments with two equalization methods ZF and MMSE. The former is $\mathit{g}_{\text{ZF},lk}(i)=\frac{\beta^{*}_{lk}(i)}{\left|\beta_{lk}(i)\right|^2}$ and the latter is $\mathit{g}_{\text{mmse},lk}(i)=\frac{\beta^{*}_{lk}(i)}{\sigma_{i,lk}^{2}+\sigma^{2}_{n,lk}+\left|\beta_{lk}(i)\right|^2}$. The results have been reported by taking average over 100 independent experiments.

\begin{figure}[htb]
     \centering
    \subfigure{\includegraphics[width=0.47\linewidth]{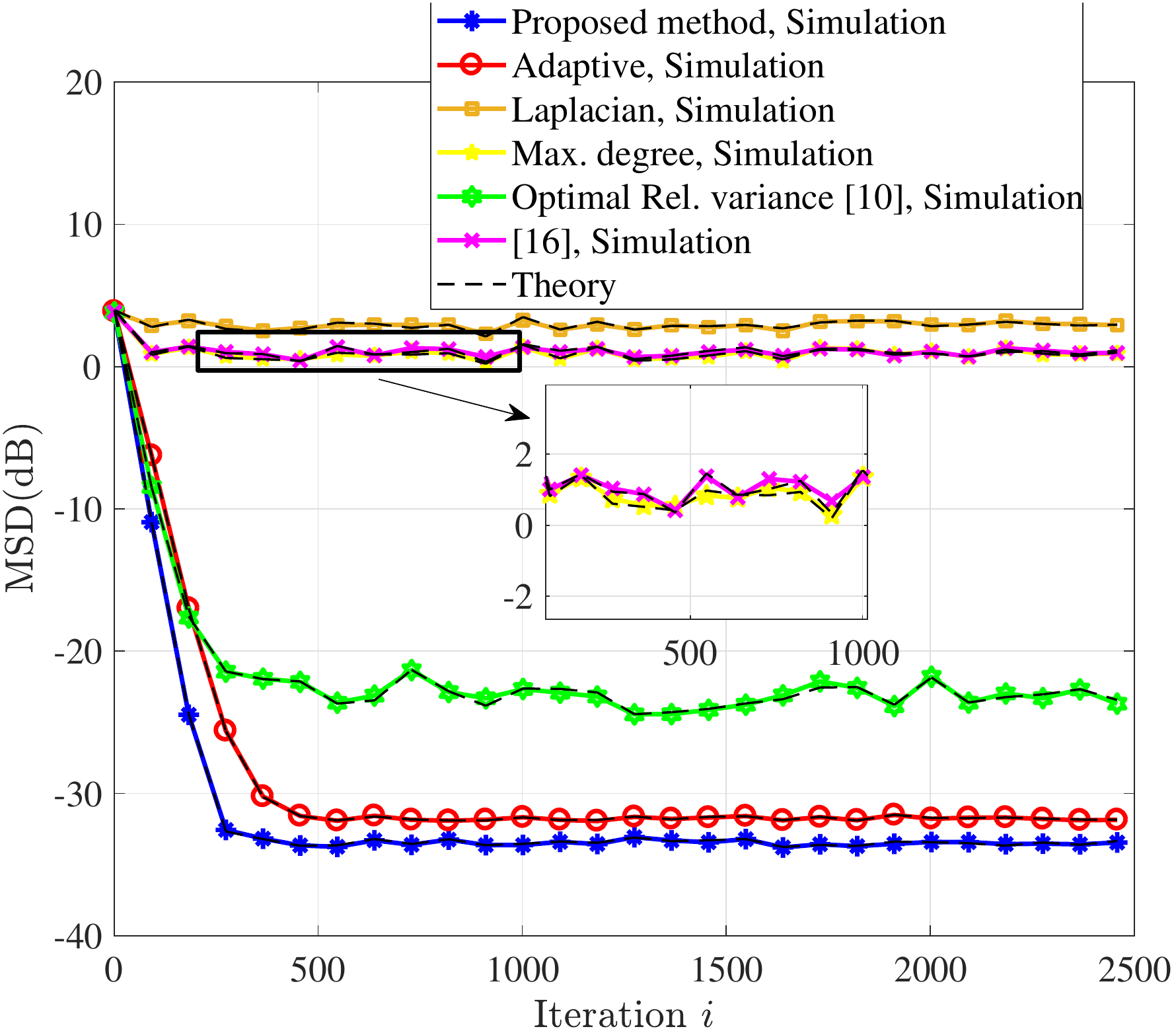}}
    \hspace*{-2mm}
\subfigure{\includegraphics[width=0.51\linewidth]{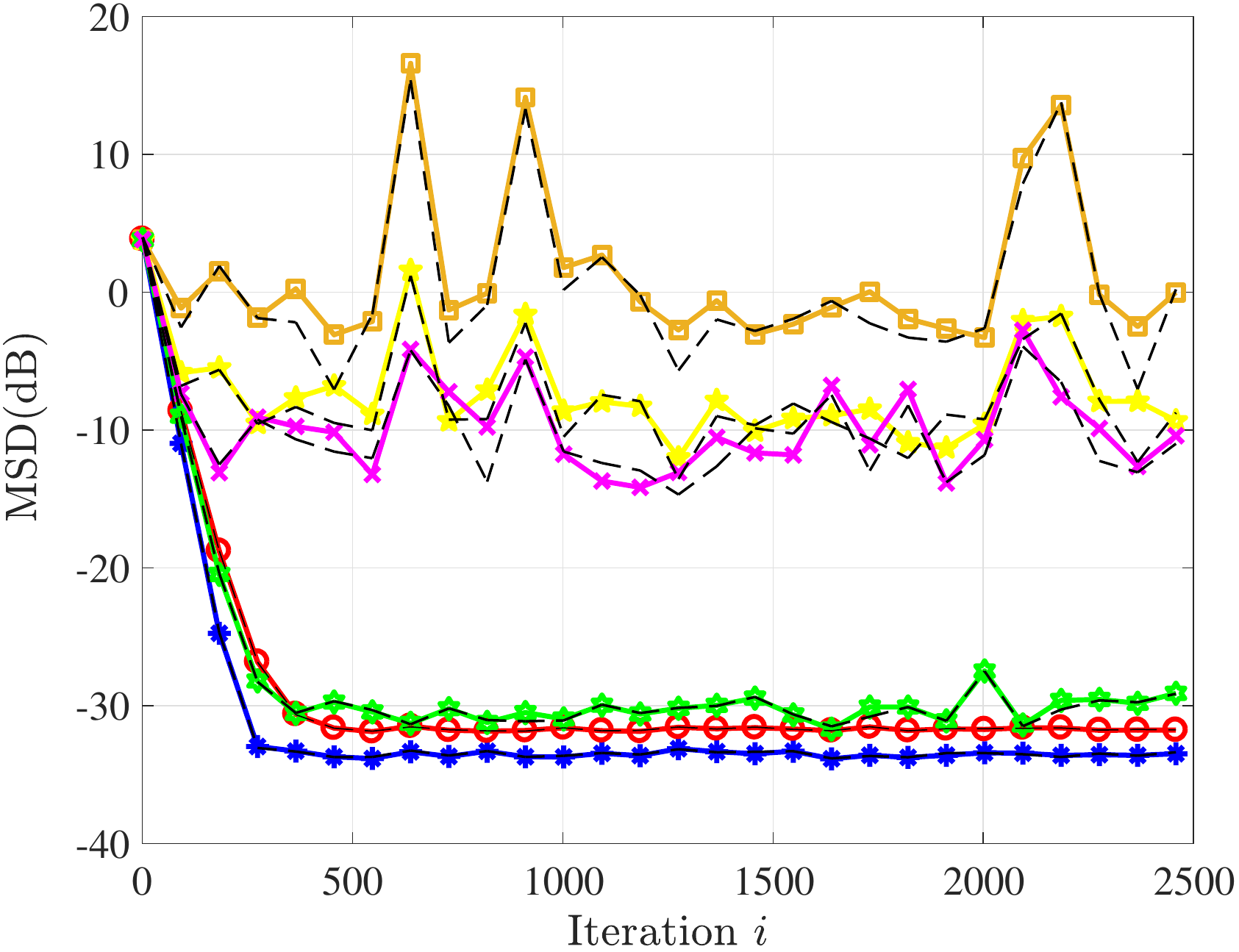}}
    \caption{\footnotesize Network MSD versus different iterations for MMSE (left) and ZF (right) equalizers. }
     \label{fig2}
\end{figure}

\begin{figure}[t]
   
    \centering
    \subfigure{\includegraphics[width=0.48\linewidth]{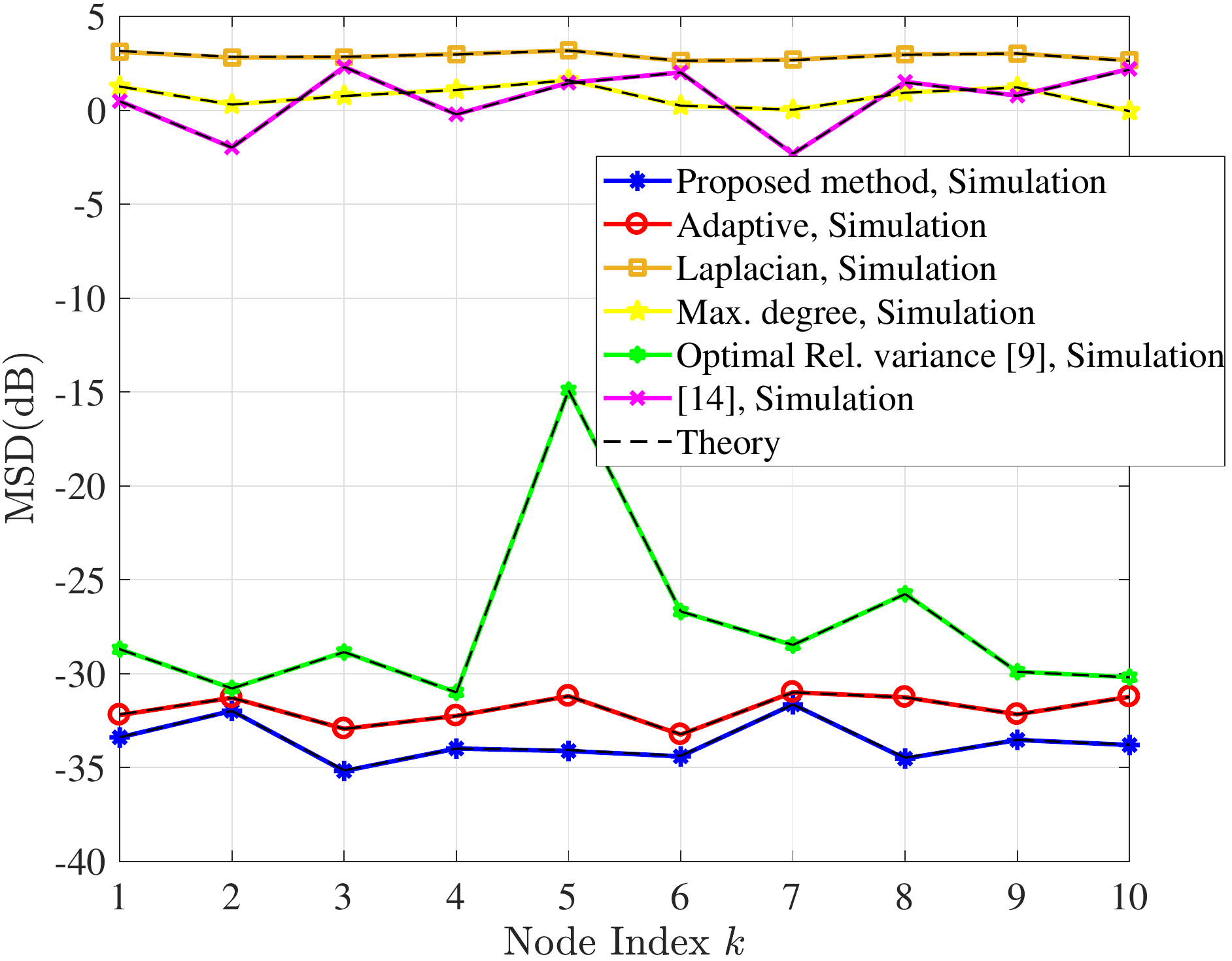}}
    \hspace*{-1mm}
    \subfigure{\includegraphics[width=0.48\linewidth]{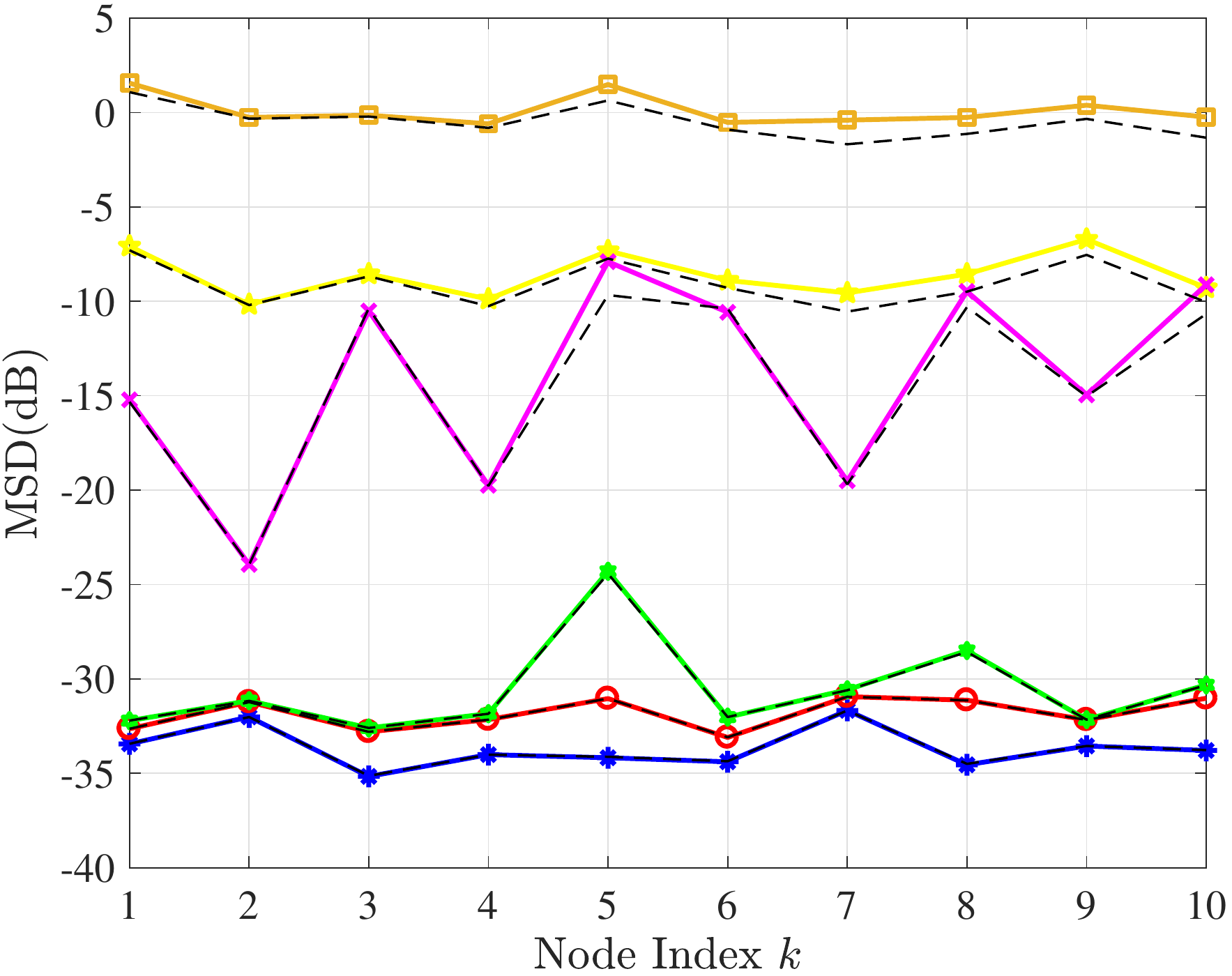}}
     
     \caption{\footnotesize Steady-state network MSD versus node index $k$ for MMSE (left) and ZF (right) equalizers. }
     \label{fig3}
\end{figure}

In these simulations, we compare the performance of the proposed optimal and adaptive combination schemes (equations (\ref{eq_21}) and (\ref{eq_25})) with Maximum Degree, Laplacian, Optimal Relative Variance (the adaptive version of these schemes are used, i.e., the combination weights will be updated if there is a change in the topology of network), and the scheme proposed in \cite{ref_12}. To this end, the network MSD verses iterations and node index are illustrated in Figure \ref{fig2} and Figure \ref{fig3}, respectively. Note that the network MSD has been calculated using $\eta(i)=\frac{1}{K}\sum_{k=1}^{K}\eta_{k}(i)$. In addition, Figure \ref{fig3} demonstrates the steady-state network MSD. The theoretical results (\ref{eq_17}) are also plotted in the same figures. 


As the figures show our proposed combination rules outperform the existing methods. Figure \ref{fig2} indicates that applying MMSE equalizer instead of ZF equalizer does not improve the performance of the proposed schemes. However, it results in better performance in steady-state for the case of using Maximum Degree, Laplacian, and the method proposed in \cite{ref_12}. Furthermore, we observe a very good match between theoretical and experimental findings. It is worth mentioning that there is a slight difference between the optimal and adaptive proposed combination which is for the approximation that is used to compute (\ref{eq_25}). 
\section{Conclusion}
In this letter, we extend ATC DLMS algorithm over wireless networks with fading to a more piratical scenario in which the inter-node-interference among nodes was considered. We computed the network error vector and then find such conditions under which the algorithm converges. Our findings show that channel coefficients can cause the algorithm to be a biased estimator unless they are compensated by ZF equalizer. In addition, we propose an optimal combination method through solving an optimization problem built on an upper bound of network MSD. Besides, we proposed the adaptive version of that, which are superior to the existing methods. In addition, we simulate the proposed combination methods for both ZF and MMSE equalizers. The results indicated that our proposed schemes give the same performance for both cases, but using MMSE equalizer leads to achieving better performance in combination rules such as Maximum degree, Laplacian, and the method proposed in \cite{ref_12}.

\bibliographystyle{IEEEtran}
\bibliography{DLMS_INI}

\end{document}